\documentstyle[epsf,preprint,prb,aps,amsfonts,amssymb]{revtex}
\tightenlines
\begin{document}

{
\draft

\title{The strain energy and Young's modulus of single-wall 
carbon nanotubes calculated from the electronic energy-band theory}
\author{ Zhou Xin$^{1}$, Zhou Jianjun$^{1}$ and Ou-Yang 
Zhong-can$^{1,2}$}
\address{$^{1}$Institute of Theoretical Physics, The Chinese 
Academy of Sciences, P. O. Box 2735, Beijing 100080, China
\\ $^{2}$Center for Advanced Study, Tsinghua University, Beijing 100084,
 China}
\date{\today}

\maketitle

\begin{abstract}
 The strain energies in straight and bent single-walled carbon nanotubes
 (SWNTs) are calculated by taking account of the total energy of all 
 the occupied band electrons. The obtained results are in good 
 agreement with previous theoretical studies and experimental observations.   
 The Young's modulus and the effective wall 
 thickness of SWNT are obtained from the bending strain energies of SWNTs with 
 various cross-sectional radii. The repulsion potential between
 ions contributes the main part of the Young's modulus of SWNT. 
 The wall thickness of SWNT comes completely from the overlap of 
 electronic orbits, and is approximately of the extension of 
 $\pi$ orbit of carbon atom. Both the Young's modulus 
 and the wall thickness are independent of the radius and the helicity 
 of SWNT, and insensitive to the fitting parameters. 
 The results show that continuum elasticity theory can serve well 
 to describe the mechanical properties of SWNTs. 
 
\end{abstract}

}


Since their discovery in 1991\cite{Ijima}, Carbon nanotubes 
(CNTs) have invoked considerable
interest\cite{Ptody} in the last decade . There are many works 
on both the theoretical \cite{Hamada,Saito,White} 
and the experimental \cite{Wildoer} studies about the electronic 
structure of CNTs, and many exciting and
novel properties have been discovered. For example, it was found that
the insulating, semi-metallic, or metallic behavior depends  
upon the radius and the helicity of CNTs \cite{Hamada}. On the 
thermal and the mechanical properties, the tubes are 
 significantly stiffer than any material presently known \cite{Dresselhaus}. 
 To understand these many intriguing properties, many groups 
 have calculated the strain energy  
 \cite{Daniel,Robertson,Adams,Lenosky,Tersoff,Ouyang} and the Young's 
 modulus \cite{Yakobson,Lu,Hernandez} of single-wall carbon 
 nanotubes (SWNTs). Among these calculations, many  
 depend on the choice of 
 an empirical potential between the carbon atoms, such as  
 Tersoff-Brenner potential \cite{Jtersoff}. 
 Lenosky {\it et al.} \cite{Lenosky} employed an 
 empirical model with three parameters reducible to a 
 continuous model with two elastic moduli\cite{Ouyang}. 
 They showed that
 the continuum elasticity model serves well to describe
 the deformation of multi-wall carbon nanotubes (MWNTs).   
 Resent theoretical studies on the Young's moduli of SWNTs 
 \cite{Yakobson,Lu,Hernandez} show some discrepancies coming from 
 the adoption of  different empirical potentials 
 and different relations in the continuum elasticity theory (CET), 
 especially, the different values of the effective wall 
 thickness of SWNT. How to calculate 
 the Young's modulus of SWNT is still an open question.  
   
 Here we present a simple method for the computation of the 
strain energy of straight SWNTs directly from the electronic
band structure without introducing any empirical potential. 
 This method had also been extended to calculate the strain 
 energy of bent tubes. It is found that the wall thickness of SWNTs
 can be calculated simply from the band electrons, and the Young's
 modulus by consideration of both the repulsion 
 energy between ions and the bond-length dependencies of the 
 electronic energy. Our results show that CET can
 well describe the bending of SWNTs and that both the Young's modulus 
 and the effective wall thickness are independent of the 
 radius and the helicity of the tubes, and insensitive to the
 fitting parameters. We obtained the Young's modulus of SWNT
 about $5$ ${\rm TPa}$, $5$ times larger than the value of 
 MWNT or graphite bulk samples, and the effective wall thickness
 about $0.7$ ${\rm \AA}$, the size of carbon atom.

Generally, the total energy of the carbon system is given by the 
sum \cite{Menon,Yu}:   
\begin{eqnarray}
E_{total}=E_{el} + E_{rep},
\end{eqnarray}
where $E_{el}$ is the sum of the energy of band electrons of the 
occupied states and $E_{rep}$ is given by a repulsive pair 
potential depending only on the distance between two carbon atoms. 
They are given by
\begin{equation}
E_{el}=\sum_{occ}E_k,
\end{equation}
and 
\begin{equation}
E_{rep}=\sum_{i} \sum_{j>i} \phi(r_{ij}),  
\end{equation}
, respectively. 
Since $\phi(r)$ is a short-range potential \cite{Menon}, only 
interaction between neighbor atoms needs to be considered. 
 On account of the relaxation effect \cite{Hamada,Saito,White}, the 
 bond length of SWNT is slightly larger than 
 that of graphite ($r_0=1.42$ ${\rm \AA}$). However,  
 even in $C_{60}$, for which the relaxation effect is significant 
 on account of its small radius, 
 calculations show that the energy contribution of 
 the bond-relaxation can still be safely ignored \cite{Lenosky}. 
 The total energy can now be rewritten as:
 \begin{equation}
     E_{total}=\frac{1}{2} {\sum} {\cal E}_{0} \cdot ({\delta} r_{ij})^2 + E_{ang},  
 \label {energy}
 \end{equation}     
 where the first term on the right hand side of $E_{total}$ is the 
 sum of the repulsion energy between ions and the electronic energy 
 contribution of the bond length change with ${\delta} r_{ij}$ as 
 the change of the distance between 
 the $i$th and the $j$th atoms in SWNT from that in graphene.  
 The second term is the electronic energy contribution 
 of the angular change of the bond, when rolling from graphene to 
 SWNT. The positions of the atoms of 
 straight SWNTs are located on the cylindrical surface of the 
 tube when the relaxation effect of the bonds is neglected.  
 ${\delta} r_{ij}$ is proportional to ${\rho}^{-2}$, 
 where $\rho$ is the cross-sectional radius of SWNT, and  
  the first term of Eq. (\ref {energy}) can be ignored, since it 
 is of $\rho ^{-4}$ order. Therefore, the strain energy of straight 
 SWNTs comes from the curvature-induced electronic energy 
 change, and can be obtained by taking account of the electronic
 energy of all the occupied bands. 
  
In order to calculate the electronic energy bands 
of SWNTs, we use a simple nearest-neighbor tight-binding (TB) model. 
This model contains nine TB 
parameters of graphite: Four hopping  including, 
$V_{ss\sigma}=-6.679$, 
$V_{sp\sigma}=-5.580$, $V_{pp\sigma}=5.037$, $V_{pp\pi}=-3.033$
in unit of ${\rm eV}$; four overlapping integration including, 
$S_{ss\sigma}=0.212$, $S_{sp\sigma}=0.102$, $S_{pp\sigma}=-0.146$, 
$S_{pp\pi}=0.129$ and  an energy difference between the $2s$ orbit
and the $2p$ orbit of the carbon atoms 
$\Delta {\cal E}$=$({\cal E}_{2s}-{\cal E}_{2p})=-8.868$ ${\rm eV}$.   
\cite{Saito} The model has been used
widely for the calculation of the electronic properties of both graphenes
and SWNTs. In general, these TB parameters depend upon the 
bond-length in the way \cite{Menon},  
\begin{equation} 
V_{\lambda {\lambda}^{'} \mu}(r)=
V_{\lambda {\lambda}^{'} \mu}(r_0) \cdot \exp(-\gamma (r-r_0)).
\label{TBparameter}
\end{equation}
However, in case of straight SWNTs, the ${\rho}^{-2}$ order 
dependence of the strain energy will not be affected even if we
 ignore simultaneously these dependencies
 and the repulsion energy. 

 With the notation used by White {\it et al.} \cite{White}, each SWNT
 is indexed by a pair of integers $(n_1,n_2)$ corresponding to the lattice 
 vector ${\vec R}$=$n_{1} {\vec a}_{1}+n_{2} {\vec a}_2$ on the graphene,
 where ${\vec a}_1$, ${\vec a}_2$ are the unit cell vectors of 
 the graphene. The tube structure is obtained by a rotation
 operation ${\cal C}_N$ and a screw operation 
 ${\cal S}(h, \alpha)$. The operation ${\cal C}_N$ is a rotation 
of $\frac{2 \pi}{N}$ about the axis, where $N$ is the largest common 
factor of $n_1$ and $n_2$. The ${\cal S}(h, \alpha)$ operation is 
 a rotation of an angle 
 $\alpha$ about the axis of SWNT in conjunction with a translation of
$h$ units along the axis, which both $h$ and $\alpha$ depending  
on the tube parameters~\cite{White}. Let $[m,l]$ denote a primitive 
unit cell in 
 the tube generated by mapping the $[0,0]$ cell to the 
 surface of the cylinder first and then
 translating and rotating this cell by $l$ applications of the 
 rotational operator ${\cal C}_N$ followed by $m$ applications of 
 ${\cal S}(h, \alpha)$.
Because ${\cal S}(h, \alpha)$ and ${\cal C}_{N}$ 
commute with each other, we can generalize the Bloch sums, 
and obtain the Hamiltonian matrix:  
\begin{eqnarray}
&&{\cal H}_{ij}^{AA}(k,n)={\cal H}_{ij}^{BB}(k,n)=
{\cal E}_i {\delta}_{ij}, \nonumber \\
&&{\cal H}_{ji}^{BA}(k,n)=({\cal H}_{ij}^{AB}(k,n))^{*}, \nonumber \\
&&{\cal H}_{ij}^{AB}(k,n)=\sum_{r} \exp[\frac{2 n i\pi}{N} \Delta l(r)
+i k \Delta m(r)]{\cal V}^{AB}_{ij}(r), 
\end{eqnarray}
where $[\Delta m(r), \Delta l(r)]$ $(r=1, 2, 3)$   
are the cell indices of the primitive unit cells located by  
the three nearest-neighbor atoms $B$ of atom $A$ in the tube, 
$A$ and $B$ are two independent carbon atoms in a primitive unit cell
of SWNT. Let $n=0, 1, \ldots, N-1$ represent the $N$ sub-Brillouin 
Zones, $k$ be a one-dimensional wave vector. ${\cal E}_1 
={\cal E}_{2s}$, ${\cal E}_i={\cal E}_{2p}$ for $i=2, 3, 4$. 
Taking the $2p$ wave function as a vector and $2s$ wave 
function a scalar, one can easily obtain:
\begin{eqnarray}
{\cal V}^{AB}_{p_i,p_j}(r)&&=(\hat e_{A_i} \cdot \hat e_{B_{j}(r)}) 
V_{pp\pi} \nonumber\\
&&-(\hat e_{A_i} \cdot \hat u_r) 
(\hat e_{B_{j}(r)} \cdot \hat u_r)(V_{pp\pi}-V_{pp\sigma}),\nonumber \\
{\cal V}^{AB}_{s,p_i}(r)&&=
(\hat e_{B_i}(r) \cdot \hat u_r) V_{sp\sigma}, \nonumber \\
{\cal V}^{AB}_{p_i,s}(r)&&=
-(\hat e_{A_i} \cdot \hat u_r) V_{sp\sigma}, \nonumber \\
{\cal V}^{AB}_{s,s}(r)&&=V_{ss\sigma}.
\label{hamiltonian}
\end{eqnarray}
where ${\hat u}_r (r=1, 2, 3)$ be the unit vector from the atom $A$ 
to its three neighboring atoms $B$. $\hat e_{A_i}$ and 
$\hat e_{B_j}(r)$ are the unit vector of the 
$2p_i$ wave function of atom $A$ and the unit vector of the 2$p_j$ 
wave function of atom $B$, respectively. 
The overlapping integration matrix ${\cal S}(k,n)$ has the 
 same form as the Hamiltonian matrix, with four overlapping
 integration parameters to replace the four hopping integration parameters,
 and with unit to replace the energy of $2s$ and $2p$ wave
 function ${\cal E}_{2s}$ and ${\cal E}_{2p}$. 
Thus we obtain an 8 $\times$ 8 Hamiltonian matrix ${\cal H}(k,n)$ 
and an overlapping integration
matrix ${\cal S}(k,n)$. By solving the secular equation 
${\cal H}(k,n) {\cal C}_{i}(k,n)$=$E_{i} {\cal S}(k,n) {\cal C}_{i}(k,n)$, 
we can calculate the electronic energy band $E_{i}$ of SWNTs.

Taking account of the total energy of all the occupied band 
electrons in SWNTs relative to that in graphene, 
we have calculated the strain energy $E_s$ of the straight SWNTs.    
With the possible bond length dependence 
of the TB parameters being neglected, and with the real bond length 
of SWNT $|{\vec u}_r|$, where ${\vec u}_r$ represents vectors between
the nearest neighbor atoms in the tube~\cite{Zhou1}, we have  
 calculated the direction cosines 
${\hat e}_i \cdot {\hat u}_r$ of Eq. (\ref{hamiltonian}).   
 Fig. 1 shows that 
$E_{s}$ depends only on the radius $\rho$ of the tubes. 
The characteristic behavior $E_{s}$=${\cal C}/{\rho}^2$ is 
found with ${\cal C} \approx 1.44$ ${\rm eV {\AA}^2/ atom}$,  
 in good agreement with previous calculated value 
1.34 \cite{Tersoff} or
1.53 ${\rm eV {\AA}^2/ atom}$ \cite{Ouyang}, and  excellently 
close to the value of 1.57 extracted from the measured phonon 
spectrum of graphite \cite{Nicklow}.

Recently, $``$curved SWNTs'' and $``$torus-like SWNTs'' have been 
found \cite{Liu}. They still have the $sp^2$ bond structure, but
it is predicted to have pentagon-heptagon defects
 \cite{Lenosky}. In $``$curved SWNTs'', the 
bond-length is nearly the same as 
that in graphite sheet, since the distortion that is created by 
the bending nature of the curved tube is topologically relaxed 
by the inclusion of fivefold and sevenfold rings. However, 
the application of an external force moment at the two ends of 
the tube gives a different deformation. 
The hexagonal structure of the tube will not change until it reaches 
a critical bending curvature \cite{sjima}. The tube undergoes only a 
simple compression on the inner side, and a stretching on the 
outer side. In the following discussion, the 
word $``$curved SWNT'' refers to the growthed SWNT with  
pentagon-heptagon defects, and the word $``$bent SWNT" refers to 
SWNT that bends with outer-stretching and inner-compressing 
deformations under external force moments applied to the two 
ends of the tube. Using an 
empirical model employed by Lenosky {\it et al.}\cite{Lenosky}, 
Ou-Yang {\it et al.} \cite{Ouyang} have developed a macroscopic 
continuous elastic model to calculate for $``$curved SWNT''. 
In their work,
the strain energy of $``$curved SWNT'' come from the angular
change of the bonds or the curvature of the tubes. However, in 
the case of 
$``$bent SWNT'', the bond length effect will contribute the main
part of the strain energy.   
 In what follows, we will treat only the latter case.

The $``$bent SWNT'' surface can be described by \cite{Ouyang}
\begin{eqnarray}
{\vec Y}(s,\phi)={\vec r}(s) + \rho [{\vec N}(s) \cos \phi +{\vec b}(s) \sin \phi]
\end{eqnarray}
where ${\vec r}(s)$ is the position vector of the axis, and 
$0<s \le l$, is the arc-length parameter along the bent SWNT axis. 
$0<\phi \le 2 \pi$. ${\vec N}(s)$ and ${\vec b}(s)$ 
are unit normal and 
unit binormal vector of ${\vec r}(s)$, respectively. 
The position of each carbon atom is described by two parameters,
$s$ and $\phi$. The two operations ${\cal C}_N$ and ${\cal S}(h, \alpha)$
can still be used to determine the positions of atoms in the 
 SWNT. Therefore, a translation of $h$ units along the 
axis of SWNT means an addition of $h$ to $s$, and a rotation of  
$\alpha$ about the axis means an addition of $\alpha$ to $\phi$. 
However, because the rotational symmetry about the axis of the 
bent SWNT is broken, ${\cal C}_N$ and ${\cal S}(h, \alpha)$ are 
not symmetry operations. It is necessary to generalize the Bloch 
sums in the crystal
unit cell containing ${\cal M} \times N$ primitive unit cells of 
SWNT. Here $\cal M$ is the length of the cell along the axis 
direction( unit is $h$ ).   
${\cal M}=2 ( {n_1}^2+{n_2}^2+n_1 n_2 )/ N^2$, for $n_1-n_2$ 
 not a multiple of $3N$, 
and ${\cal M}=2( {n_1}^2+{n_2}^2+n_1 n_2 )/(3 N^2)$,
 for $n_1-n_2$ a multiple of $3N$ \cite{Jishi}. 
We calculate only SWNT with constant 
radius of curvature $R$. 
It is not principal difficult to extend the present 
treatment to general bent SWNTs. 
In bent SWNT, ${\cal V}^{AB}_{ij}(r)$ depends on the 
position of the atom $A$, and will be written as 
${\cal V}^{AB}_{ij}(l,m;r)$, where 
$[m,{l}]$ are indices of the primitive unit cell of  
the atom $A$. 
When SWNT is bent to a different direction ${\vec N}(s)$, 
${\cal V}^{AB}_{ij}(r)$ will be different, but the
Hamiltonian matrix elements are almost independent of the bending 
direction. We have found that the anisotropy effect 
is very small. Similar to the case of the straight SWNT, it is 
easy to obtain the 
$(8 \cdot {\cal M} \cdot N) \times (8 \cdot {\cal M} \cdot N)$ 
matrices of the  
Hamiltonian and the overlapping integration.
   
 Since the change of bond length $\delta r$ in bent SWNT is 
 proportional to $\frac{\rho}{R}$, 
 the energy contribution of the bond stretching and bond compressing 
 will be of the order of $\frac{1}{R^2}$. It is necessary to 
 calculate
 both the electronic energy $E_{el}$ and the repulsion energy 
 $E_{rep}$ between the ions. In order to fit the force constant of 
 graphite \cite{Rajishi}, 
 we take $\gamma=1.024$ ${\rm \AA^{-1}}$, 
 ${\phi}^{'}=\frac{\partial \phi}{\partial r}=-13.63 \gamma$ 
 ${\rm eV/{\AA}}$
 and ${\phi}^{''}=60.4$ ${\rm eV/ {\AA}^2}$. 
 With these parameters, we arrive correctly the second derivative 
 of the stretching energy $E_c$ of SWNTs be
 $D=\frac{\partial^2 E_c}{{\partial \epsilon}^2}=58.5$ ${\rm eV}$ 
 and the Poisson ratio be $\sigma=0.24$, where $\epsilon$ is the relative 
 compression along the axis of SWNT.  
 
Fig. 2${\bf (a)}$ shows the strain energy $E_b$ per atom of the (5,5) 
SWNT as a function of the bending radius $R$. 
 The data follow quite well with the expected behavior   
 $E_{b} = E_s+{\lambda}/R^2$. A least square fit to the data yields
 a value of $\lambda \approx 173$ ${\rm eV {\AA}^2/ atom}$. 
 Previous studies on $``$curved SWNTs'' give a simple  
 formula can be given\cite{Ouyang}, 
 \begin{equation}
 E_b=\frac{{\cal C} R}{{\rho}^2 \sqrt{R^2-{\rho}^2}}
 \approx \frac{{\cal C}}{{\rho}^2}+\frac{{\cal C}}{2 R^2}. 
 \end{equation}
 In comparing with our results, we find that the value of $\lambda$ 
 for $``$curved 
 SWNT'' is equal to ${\cal C}/2$, only $0.7$ ${\rm eV {\AA}^2/ atom}$. 
 It implies that the strain energy of pentagon-heptagon defects is
 far less than the strain energy of the stretching and 
 compressing of the bond length. 
 The experimental fact that the deformations of $``$bent SWNT" are
 change of bond length rather than the pentagon-heptagon defect 
 reveals that
  there is a high potential barrier between the two 
  deformations to prevent the change of the hexagonal structure 
  under the addition of a moment of external force at the two
  end of SWNT. 

By CET, we can calculate the Young's modulus $Y$ of SWNT from three
different strain energies. They are the rolling  
energy $E_s$, the compressing or stretching energy $E_c$ and the bending 
strain energy ${\Delta} E_b$. The three energies are given by
\begin{eqnarray}
 &&E_s=\frac{{\cal C}}{{\rho}^2}, \\
 &&E_c=\frac{1}{2} D {\epsilon}^2, \\
 &&{\Delta} E_b=\frac{\lambda}{R^2},
 \end{eqnarray} 
 where $\epsilon$ is the relative stretch or compression along the 
 axis of SWNTs, and $D$ is the second derivative of $E_c$. 
 The three quantities ${\cal C}$, $D$ and $\lambda$ are 
 given by
\begin{eqnarray}
&&  {\cal C}=\frac{\Omega}{24 (1-{\sigma}^2)} Y \cdot b^3,  \label {rolling} \\
&&  D=\Omega \cdot Y \cdot b,  \label {stretching} \\
&&  \lambda=\frac{\Omega}{4} Y \cdot b ({\rho}^2+b^2/4),  \label {bending} 
\end{eqnarray}
respectively. Where $\Omega=2.62$ ${\rm {\AA}^2/atom}$ 
is the occupied area per carbon atom in SWNTs, $b$ is the effective 
wall thickness of SWNT. 
Previous calculations \cite{Daniel,Yakobson,Lu,Hernandez} indicate 
that the value of $D$ of SWNTs is about $58$ ${\rm eV/atom}$, same as
that of graphites. 
 However, since the wall thickness is not well defined 
in single-layered structure, various values of $b$ are used in the
studies, thus the obtained Y are quite different.  
 Lu \cite{Lu} and Hern\'{a}ndez {\it et al.} \cite{Hernandez}
 took the interwall distance of graphite ($3.4$ ${\rm \AA}$) as 
 the thickness, and obtained the average Young's modulus of SWNT 
 about $1$ ${\rm TPa}$, in consistency with the corresponding 
 measurement in multiwall nanotubes \cite{Wong} and bulk graphite 
 samples. 
 But the average value of $Y$ can not describe all kinds of 
 deformations of SWNTs, such as the rolling of graphene and the bending
 of SWNT, though it can describe the stretching and compressing
 deformation along the axis direction of SWNT. 
 Yakobson {\it et al.} \cite{Yakobson} have given  
 $Y=5.5$ ${\rm TPa}$ and $b=0.66$ ${\rm \AA}$ by using 
 the rolling energy formula of graphite sheet  
 [ Eq. (\ref {stretching})] and the stretching energy formula of 
 graphene or SWNT [Eq. (\ref {rolling})], simultaneously. The 
 obtained value of $b$ is about 
 the $\pi$ orbital extension of carbon atom, which corresponds to 
 the general fact that elasticity results from the overlapping of 
 electron cloud between 
 atoms. However, since Eq. (\ref {rolling}) describe the rolling of 
 single-layered graphene, the results given by Yakobson seems to 
 correspond to graphite sheet rather than SWNT. The Young's
 modulus and the effective wall thickness of SWNT and their dependence of
 the tube radius and helicity still remain unknown. 
   
With Eq. (\ref {bending}), from the calculation of the bending 
strain energy of SWNTs with various radius, one may simultaneously 
find $Y$ and $b$ of SWNTs. Fig. 2${\bf (b)}$ shows the relationship 
between $\lambda$ and the radius of the tube in the form 
$\lambda=b^{*} {\rho}^2+a^{*}$. In comparison with the 
 Eq. (\ref{bending}), it implies that both $Y$ and $b$ are 
 independent of the radius and helicity of SWNTs. The value of 
 $b^{*}=15.3$ ${\rm eV/atom}$ is 
 consistent with the value $D/4=14.7$ ${\rm eV/atom}$. 
 However, it is difficult to obtain the exact value of $a^{*}$, 
 because the first term of $\lambda$ is much greater than the second 
 term $a^{*}$, to introduce high errors in the our fitting of 
$a^{*}$. Carefully analysis of the strain energy of $``$bent SWNT'', 
gives only that the electronic energy from the angular change of the
bond can contribute to the ${\rho}^{0}$ order of the bending strain 
energy of SWNT. The other terms, including the
repulsion energy between ions and the electronic energy from the 
nonzero $\gamma$ effects (Eq. (\ref{TBparameter})), depend on 
${\delta} r$, hence on the radius of the tube $\rho$. For 
 wall thickness, is is unnecessary to consider the bond length 
 dependence of TB parameters and the repulsion energy. It require 
 only to calculate $\gamma = 0$ bond angular contribution
$E_{el0}$ of the electronic energy. When $\gamma = 0$, the 
$\frac{1}{R}$th order perturbation of Hamiltonian is zero, and 
$E_{el0}=\frac{{\lambda}_{el0}}{R^2}$. The ${\rho}^0$ order term of 
$\lambda$ come completely from the ${\lambda}_{el0}$ and the 
residual part of $\lambda$ affects only the value of ${\rho}^2$ 
order. The exact value of $a^{*}$ can be obtained by calculating 
$E_{el0}$. Fig. 2 {\bf (c)} shows the expected relationship 
 $E_{el0}={\lambda}_{el0}/R^2$. 
Fig. 2${\bf (d)}$ gives values of ${\lambda}_{el0}$ for 
some SWNTs. It lead to 
$a^{*}=1.05 {r_0}^2$ ${\rm eV {\AA}^2/ atom}$ by 
fitting it to $a^{*}+a_1 {\rho}+a_2 {\rho}^2$.\cite{Zhou2}  
The wall thickness of the tube is supposed to be identical to that 
of graphite, then from Eq. (\ref{rolling}) and Eq. (\ref{bending}), 
it gives $a^{*}=  
\frac{3}{2} (1-{\sigma}^2) {\cal C} \approx 1.0 {r_0}^2$ 
${\rm eV {\AA}^2/atom}$. 
Therefore, $b$ is about $0.74$ ${\rm \AA}$, and $Y$ is about
 $5.1$ ${\rm TPa}$. This shows that both $Y$ and $b$ are independent 
 of the radius and helicity of the tube, and the Young's modulus of 
 SWNT is five times greater than the average value of MWNT. 
 The obtained value of $b$ is independent of the 
 fitting parameters $\gamma$ and $\phi^{''}$, and $Y$ is also
 insensitive to these parameters.
  
In summary, our calculation shows the following results: the strain 
energy of the straight SWNT come mainly from 
 the occupied bands electrons, The obtained Young's modulus 
of SWNT is independent of the radius and the helicity and is much 
larger than the modulus of the bulk sample. The effective 
thickness of SWNT is about the size of the carbon atom, far less
 than the distance between the layers of the graphite. 
These results show that CET can well describe the deformation of 
the bent tubes.   
   
 The authors acknowledge the useful discussions in our group. We 
 would like to thank Dr. X.-J. Bi, Mr. Y.-H. Su and G.-R. Jin for 
 correcting an earlier version of the manuscript and thank Prof. 
 Y.-Z. Xie and Dr. H.-J. Zhou for correcting the manuscript in
 English. The numerical calculations
 are performed partly at ITP-Net and partly at the State Key Lab. 
 of Scientific and Engineering Computing.

\begin{figure}
\centerline{\epsfxsize=10cm \epsfbox{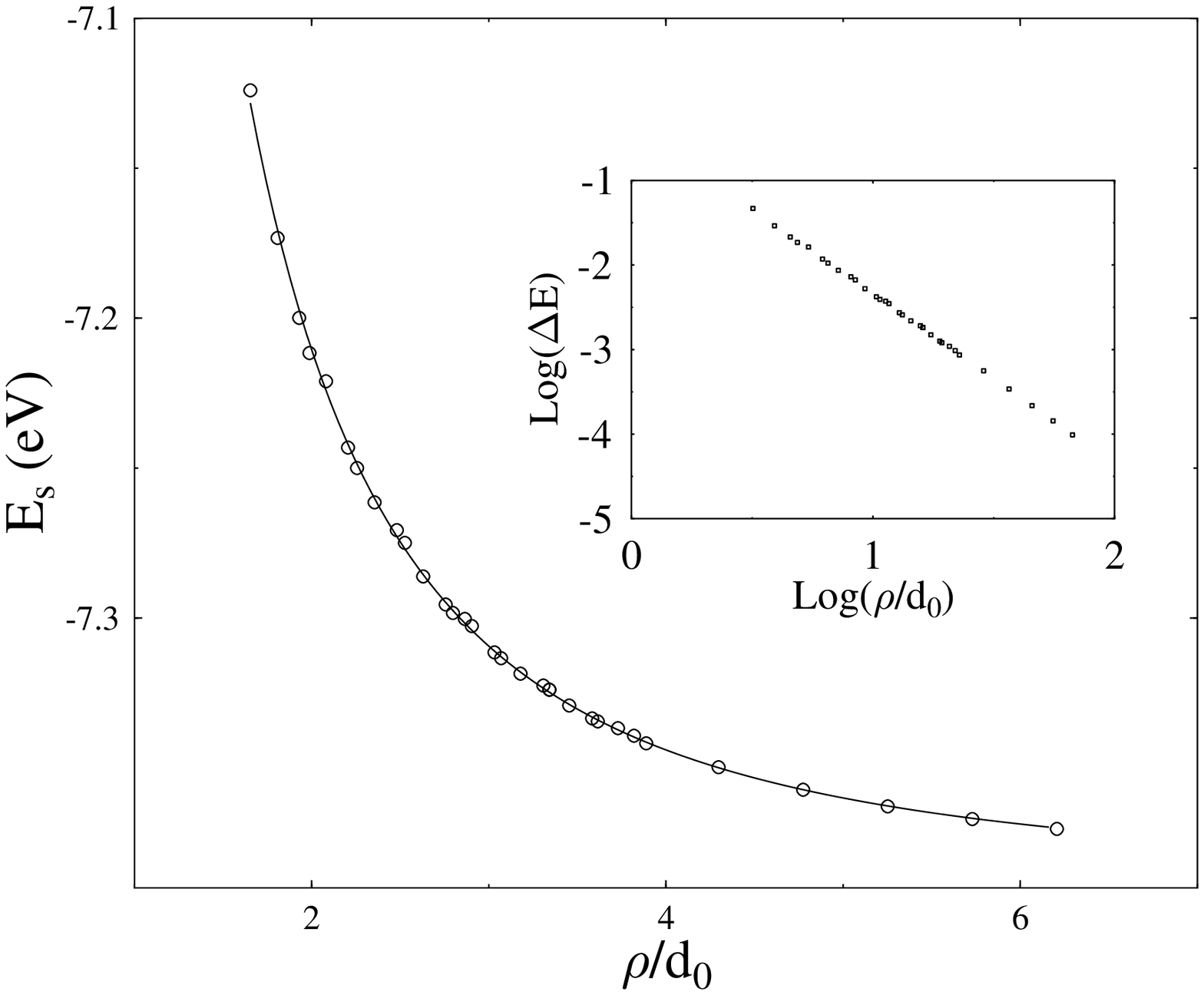}}
\caption{The strain energy per atom versus the radius of 
$(n, m)$ tubes, where $n=6 \sim 13$ and $m=0, 1, 2$ or $n$. The solid
 line corresponds to a least square fit to the 
 $\frac{\cal C}{\rho^2}$ 
behaviors. ${\cal C}=0.71 \cdot {r_0}^2 \approx 1.44$
 ${\rm eV {\AA}^2/ atom}$. ${\cal C}$ is independent on the 
 helicity of tubes. The ${\rho}^{-\beta}$ behaviors is clearly 
 shown in the inset ($\beta \approx 2.03$ ).
 Here $r_0$ is $1.42$ ${\AA}$, the cohesive energy of graphite 
 is $-7.39$ ${\rm eV}$.  
\label{fig1}}
\end{figure}


\begin{figure}
\centerline{\epsfxsize=10cm \epsfbox{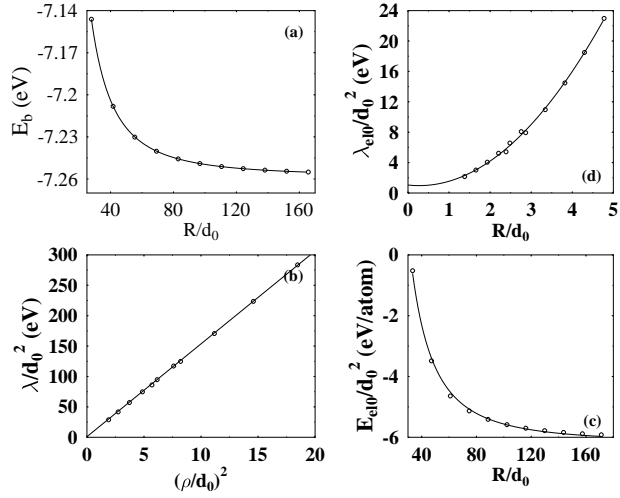}}
\caption{{\bf (a)} Strain energy per atom 
versus the bending radius R in (5,5) tube. 
The solid line is a fit to the $E_s+\lambda/R^2$, where $E_s$
is the cohesive energy of straight (5,5) tube . 
$\lambda=86.1 \times 1.42^2$ ${\rm eV {\AA}^2/ atom}$. 
{\bf (b)} the value of $\lambda$ of some (n,0) and (n,n) tubes. 
The solid line is a fit to $a^{*}+b^{*} {\rho}^2$, $b^{*} \approx 15.27$ 
${\rm eV}$. 
{\bf (c)} $E_{el0}$ versus the bending radius R 
in (5,5) tube. The solid 
line is a fit to $E_0+\lambda_{el0}/R^2$. The zero point of
$E_{el0}$ has moved. 
{\bf (d)} The value of $\lambda_{el0}$ of these tubes. The solid 
line is a fit to $a_0+a_1 \rho +a_2 {\rho}^2$, and the 
$a_0 \approx 1.05$. 
\label{fig2}}
\end{figure}


\begin{thebibliography}{99}
\bibitem{Zhou2} the $\rho$ order contribution of $\lambda_{el0}$ 
may be nonzero, although the term of the total $\lambda$ vanishes.
\end{thebibliography}
\end{document}